\documentclass{nature}

\usepackage{amsmath}
\usepackage[pdftex]{graphicx}
\usepackage{dcolumn}  
\usepackage{bm}       
\usepackage{dsfont}
\usepackage{color}
\usepackage{braket}

\newcounter{defcounter}
\usepackage{amsthm}

\title{Coherent Dynamics of Charge Carriers in $\rm \gamma$-InSe Revealed by Ultrafast Spectroscopy} 

\author{Jianwei Shen$^{1}$, Jiayu Liang$^{1}$, Qixu Zhao$^{1}$, Menghui Jia$^{1}$, Jinquan Chen$^{1}$, Haitao Sun$^{1}$, Qinghong Yuan$^{1}$, Hong-Guang Duan$^{2}$, Ajay Jha$^{3,4,5}$, Yan Yang$^{1}$ \& Zhenrong Sun$^{1}$ } 

\begin{document} 

\maketitle

\begin{affiliations}
\item State Key Laboratory of Precision Spectroscopy, School of Physics and Electronic Science, East 
China Normal University, Shanghai 200241, P.R. China 
\item Department of Physics and Institute of Modern Physics, Ningbo University, Ningbo 315211, China
\item Rosalind Franklin Institute, Harwell, Oxfordshire OX11 0QX, United Kingdom
\item Department of Pharmacology, University of Oxford, Oxford OX1 3QT, United Kingdom 
\item Research Complex at Harwell, Rutherford Appleton Laboratory, Didcot OX11 0QX, United Kingdom \\ 
\centerline{\underline{\date{\bf \today}}} 
\end{affiliations} 

\begin{abstract} 

For highly efficient ultrathin solar cells, layered indium selenide (InSe), a van der Waals solid, has shown a great promise. In this paper, we study the coherent dynamics of charge carriers generation in $\rm \gamma$-InSe single crystals. We employ ultrafast transient absorption spectroscopy to examine the dynamics of hot electrons after resonant photoexcitation. To study the effect of excess kinetic energy of electrons after creating A exciton (VB1 to CB transition), we excite the sample with broadband pulses centered at 600, 650, 700 and 750 nm, respectively. We analyze the relaxation and recombination dynamics in $\rm \gamma$-InSe by global fitting approach. Five decay associated spectra with their associated lifetimes are obtained, which have been assigned to intraband vibrational relaxation and interband recombination processes. We extract characteristic carrier thermalization times from 1 to 10 ps. To examine the coherent vibrations accompanying intraband relaxation dynamics, we analyze the kinetics by fitting to exponential functions and the obtained residuals are further processed for vibrational analysis. A few key phonon coherences are resolved and ab-initio quantum calculations reveal the nature of the associated phonons. The wavelet analysis is employed to study the time evolution of the observed coherences, which show that the low-frequency coherences last for more than 5 ps. Associated calculations reveal that the contribution of the intralayer phonon modes is the key determining factor for the scattering between free electrons and lattice. Our results provide fundamental insights into the photophysics in InSe and help to unravel their potential for high-performance optoelectronic devices.

\end{abstract} 


The van der Waals chalcogenides show a variety of different specific functions that depend on their composition and number of layers. Weak interaction of materials along the stacking direction facilitates the realization of heterostructures with different functionalities. Recent reports demonstrate possible ways for fine tuning the band gap \cite{PRL 105 136805 2010, PRB 89 235319 2014, Adv Mater 25 5714 2013}, controlling valley polarization \cite{Nat Nanotechnol 7 494 2012}, and the realization of high charge mobilities \cite{Nat Nanotechnol 6 147 2011}. Two-dimensional InSe has attracted extensive attention due to its large tunable band gap from 1.4 to 2.6 eV and high charge carrier mobility. A few-layer InSe has been synthesized via physical and chemical methods, which exhibits promising characteristics for optoelectronic application \cite{NanoLett 14 2800 2014}.  Tamalampudi {\em et al.} showed that devices based on few-layer InSe obtained by mechanical exfoliation can be used as photosensors with high photo-responsivity \cite{NanoLett 14 2800 2014}. The experimental measurement has shown the carrier mobilities reaching (or exceeding) $10^{3}$ cm$^{2}$ V$^{-1}$ s$^{-1}$ at room temperature \cite{AdvMater 26 6587 2014, NatNanotechnol 12 223 2017}, which makes it a champion material in transition-metal dichalcogenides, TMDCs. To further optimize the optoelectronics devices based on InSe, a detailed insight to the photophysics of the material is eminent.

\begin{figure}[h!]
\begin{center}
\includegraphics[width=17.0cm]{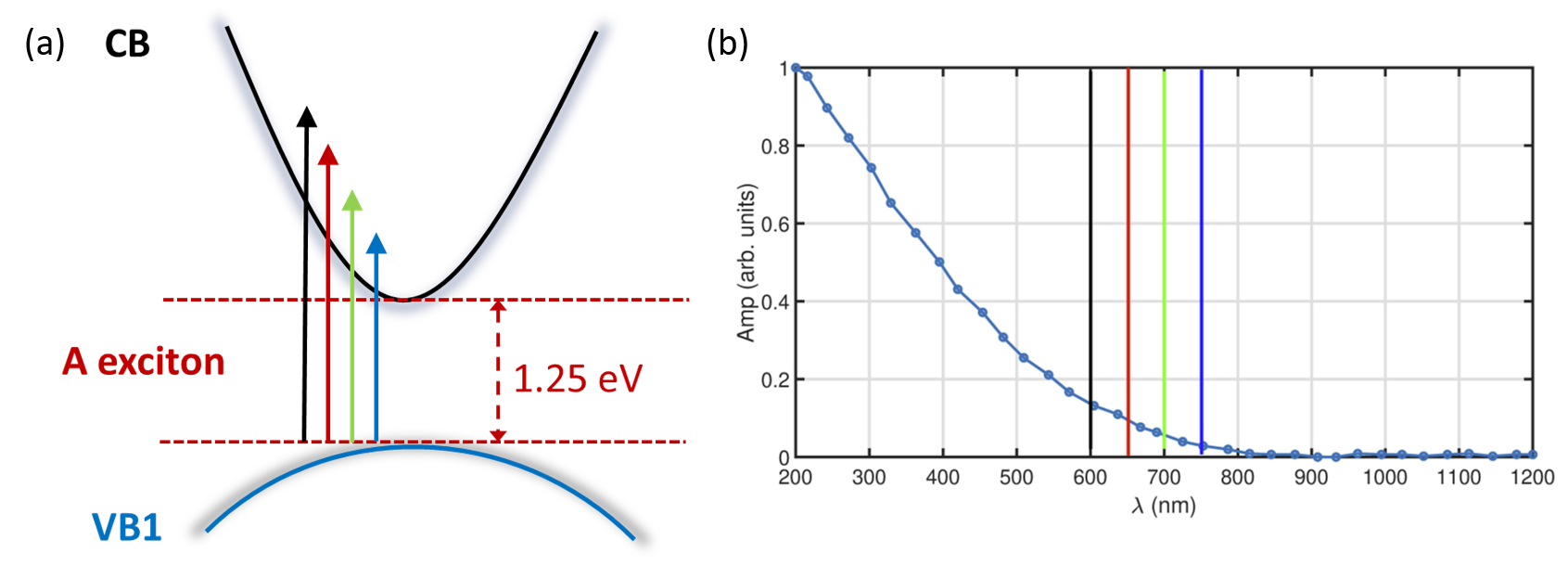} 
\caption{\label{fig:Fig1} (a) schematic showing the lowest energy electronic transitions of InSe probed in this study. Exciton A is created after transition from valence band 1 (VB1) to conduction band (CB) (bandgap is 1.25 eV i.e. 991 nm). (b) the absorption spectrum (blue dots and solid line) of $\rm \gamma$ phase InSe crystal and the center wavelengths of excitation pulses used in this study. } 
\end{center}
\end{figure}

A schematic of lowest energy electronic transition leading to A (VB1 to CB transition) exction in single-crystal InSe is depicted here in Fig.\ \ref{fig:Fig1}. To have a comprehensive understanding towards photoinduced processes in InSe after resonant excitation to generate A excitons, one needs to follow the electronic and nuclear evolution with ultrafast time-resolution. Weiss and colleagues have studied the hot carrier and surface recombination dynamics in layered InSe crystals after photoexciation to generation A and B excitions using transient reflectivity measurements. The hot carrier lifetime was found to be dependent on excited carrier density, indicating that the cooling process is caused by phonon scattering \cite{JPCL 10 493 2019}. Lu {\em et al.} have studied the photogenerated carrier relaxation dynamics using optical pump-terahertz probe measurements in InSe multilayer. The intraband relaxation showed biexponential decay rates followed by recombination in $\sim$100 ps timescale \cite{PRB 102 014314 2020}. Chen {\em et al.} employed two-photon photoemission spectroscopy to study the ultrafast electron dynamics in the material of InSe \cite{PRB97 241201 2018}. They observed that the dissipation of hot electrons occurs only via Fr{\"o}hlich coupling to optical phonons, which indicates the InSe has an excellent potential for hot-carrier optoelectronics. The subsequent studies of hot carriers in InSe show many-body screening of quasi-two-dimensional electron gas dramatically reduces the Fr{\"o}hlich scattering strength \cite{PNAS 117 21962 2020}. Desipate these extensive studies, the intrinsic mechanism for the high charge mobility and the initial phonon-coupled quantum dynamics of charge carriers in InSe is still elusive. 

In this work, we focus our attention on quantum coherent dynamics after subjecting the $\rm \gamma$-InSe single crystals to the lowest energy electronic transition, which involves generation of A exciton. We employ ultrafast transient absorption spectroscopy using broadband probe in the visible region of the spectrum to capture charge carrier relaxation dynamics after resonant photoexcitation. We selectively excite the sample within different regions of the lowest energy transition envelop using various pump photon wavelengths (600, 650, 700 and 750 nm) to generate carriers with different kinetic energies. The spectral and temporal dynamics of hot carriers have been analyzed by global fitting approach. The systematic variation in carrier relaxation pathways and lifetimes are observed. We also extract vibrational coherences from the kinetic data of ground-state bleach (GSB) band. The time-evolution of these vibrational coherences has been examined by wavelet analysis. To reveal the nature of the observed long-lived coherences, ab-initio calculations have been performed, which reveals the important contribution of intralayer phonon modes in carrier relaxation processes. Our study provides a detailed understanding towards charge carrier dynamics near the band edge in $\rm \gamma$-InSe single crystals. 

\section*{Results} 

The $\rm \gamma$-InSe sample was directly purchased from 2D semiconductor (home page: www.2dsemiconductors.com) and use it without further modification. To perform transient spectroscopic studies in transmission mode, ultrathin slices of the sample have been prepared using microtoming machine (Ultramicrotome Leica EM UC7). To check for the thickness dependence on transient dyanmics, we microtome samples with three different thicknesses, 30, 50 and 80 nm. 

\subsection{Femtosecond Transient absorption measurements}  

Prepared thin slices of the $\rm \gamma$-InSe sample are placed over quartz substrate and then mounted on a XYZ translation sample-stage. The pump/probe laser pulses are focused on the sample spot. The optical measurements have been performed after optimizing the focus sizes of pump and probe beams with minimized optical scattering. All the measurements reported in this work have been performed at room temperature. 

\begin{figure}[h!]
\begin{center}
\includegraphics[width=17.0cm]{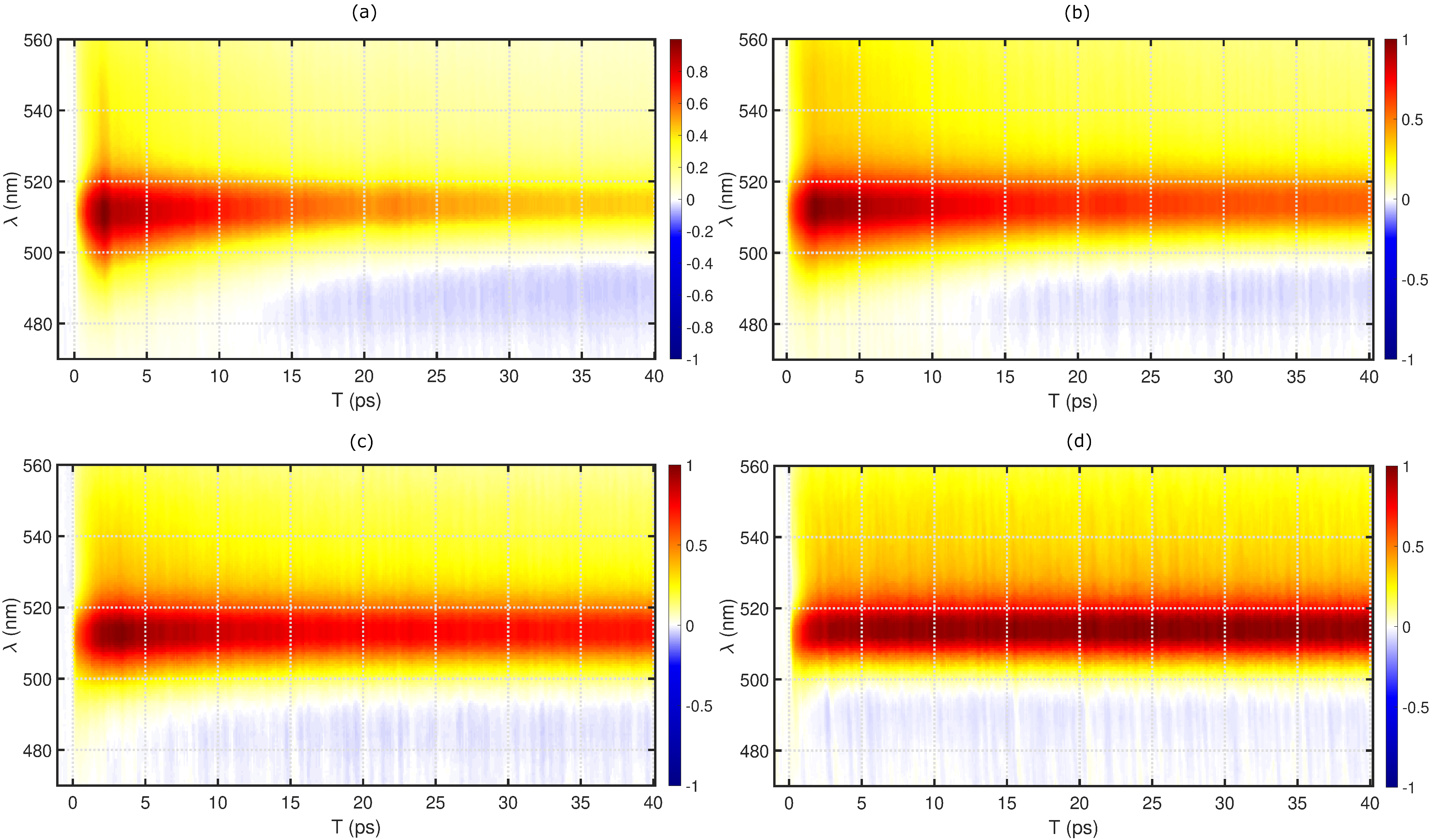}
\caption{\label{fig:Fig2} Femtosecond transient absorption (TA) measurements: TA spectra of microtomed InSe crystals obtained after excitation at 600 (a), 650 (b), 700 (c) and 750 nm (d), respectively. The positive (red) and negative (blue) amplitude indicate the GSB, SE and ESA, respectively. } 
\end{center}
\end{figure}

With the experimental condition described above, we preformed the optical measurements of InSe and show the transient absorption (TA) spectra for 50 nm thin microtomed crystals in Fig.\ \ref{fig:Fig2}. We also measure the TA data with the dependence of sample thickness and we do not observe any significant difference of results with different sample thickness. The TA data with sample thickness are shown in the Supporting Information (SI). In Fig.\ \ref{fig:Fig2}(a), transient absorption data is shown for pump excitation at 600 nm. The data shows ground state bleach feature (GSB) centered at 510 nm, which lives long. Additionally, the TA spectrum has been measured by pumping at 650 nm (shown in Fig.\ \ref{fig:Fig2}(b)), the associated broadband pulses of excitation have been shown in Fig.\ \ref{fig:Fig1}(b)). Similar to the TA data for 600 nm, it presents a positive and a negative band centered at 510 nm and 470 nm, respectively. The positive magnitude indicates the GSB and stimulated emission (SE) signals in TA spectrum, the negative one originates from excited state absorption (ESA). Compare to the GSB, the ESA signal is much weaker in the measurement. Moreover, we show the TA spectrum pumped at 700 nm in Fig.\ \ref{fig:Fig2}(c). It presents the GSB and ESA bands at 510 and 480 nm, respectively. The basic features is very similar to the TA spectrum in (b). The TA spectra pumped at 750 nm is shown in Fig.\ \ref{fig:Fig2}(d). It shows the GSB and ESA bands at 510 and 480 nm as well. However, it shows, at initial time, the fast decay of magnitude in (b) compare to the case in (d). To observe the effect of carrier densities on relaxation dynamics, we have also performed transient absorption measurements using different pump powers (shown in the SI). It can be seen that the relaxation dynamics is strongly dependent on the pump power and hence, the carrier densities. This signifies that the hot carrier cooling process occur through phonon scattering. All the data shown in Fig.\ \ref{fig:Fig2} has been measured at 80 $\mu$W.

\subsection{Global fitting results } 

\begin{figure}[h!]
\begin{center}
\includegraphics[width=17.0cm]{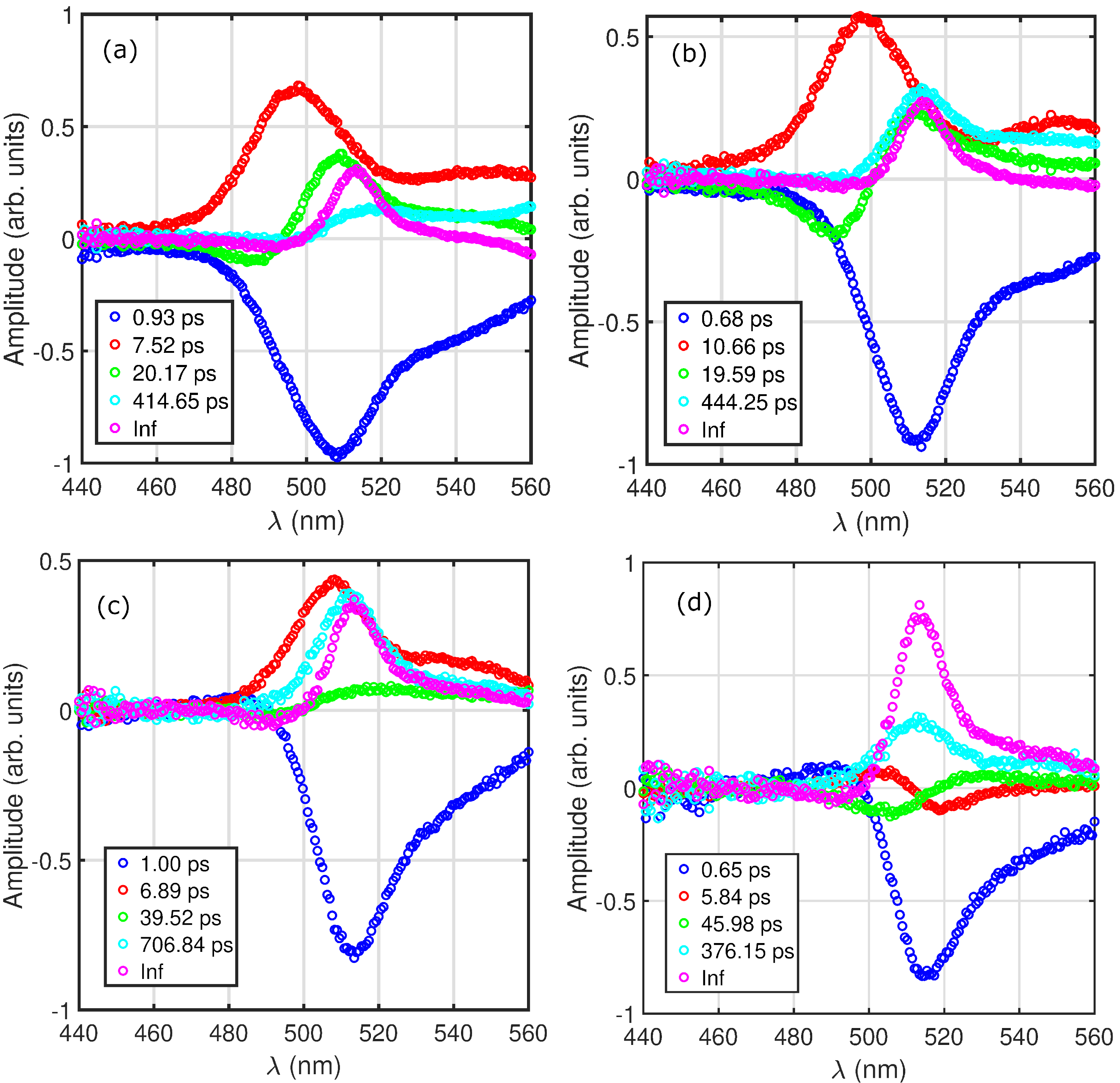}
\caption{\label{fig:Fig3} Decay associated spectra (DAS) obtained by employing global fitting approach. The DAS components with their respective lifetimes are presented for different pump excitation wavelengths: (a) 600 nm, (b) 650 nm, (c) 700 nm and (d) 750 nm, respectively. } 
\end{center}
\end{figure}

\begin{figure}[h!]
\begin{center}
\includegraphics[width=17.0cm]{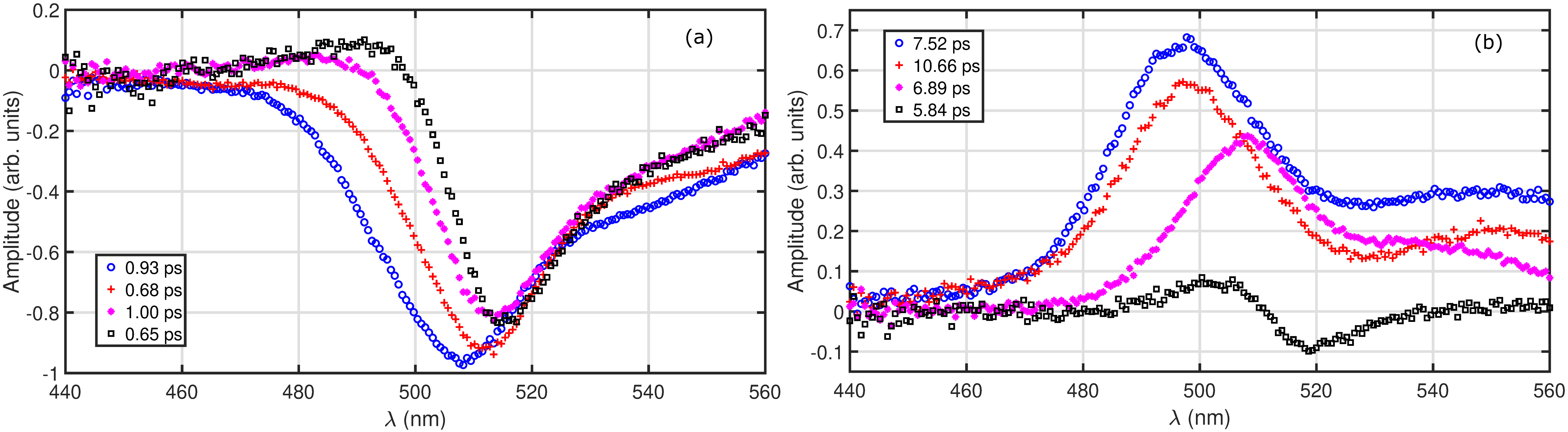}
\caption{\label{fig:Fig4} Overlay of the first and the second DAS components obtained using global fitting of the transient absorption data for different pump excitation wavelengths. The first and second components represents the carrier relaxation dynamics. The fastest components show the lifetime of 0.93, 0.68, 1.0 and 0.65 ps for the excitation of 600, 650, 700 and 750 nm in (a). The second components show the lifetime of 7.52, 10.66, 6.89 and 5.84 ps in (b). } 
\end{center}
\end{figure}

To study the time-evolved magnitude in TA, we further employ the global fitting approach to examine the kinetics of TA data. The detailed description of global fitting method has been described in the SI. In global fitting algorithm, the TA data has been fitted by multi-exponential functions to retrieve the decay components and the associated lifetimes at each probing wavelengths. By this, we obtain the decay-associated spectrum (DAS) with decay lifetime of each spectrum and we show them in Fig.\ \ref{fig:Fig3}. The DAS of pumping at 600 nm is shown in Fig.\ \ref{fig:Fig3}(a), 5 components with different lifetimes are presented. The fastest DAS shows a broadband profile and centered at 510 nm. It shows the decay lifetime of 0.93 ps and plotted as blue circles. The negative magnitude of this DAS indicates the fast growing of amplitude in TA in Fig.\ \ref{fig:Fig2}(a). Moreover, the fastest components from Fig.\ \ref{fig:Fig3}(b) to (d) show the similar spectral profiles and lifetimes of 0.68, 1.0 and 0.65 ps, respectively, which imply the generation of charge carriers in InSe. The second component shows the decay lifetime of 7.52 ps (red circles). It presents a broadband profile with positive amplitude and the spectrum is centered at 500 nm. The positive magnitude of DAS indicates the decay of magnitude in TA in Fig.\ \ref{fig:Fig2}(a). This components in Fig.\ \ref{fig:Fig3}(b) to (d) shows the lifetimes of 10.6, 6.9 and 5.8 ps, respectively. Moreover, the spectral profiles from Fig.\ \ref{fig:Fig3}(a) to (c) are roughly identical, but the amplitudes are gradually decay from (a) to (c), which imply that the contribution of the first decay component decrease with varying of the pumping wavelength. The broadband profiles in (a) to (c) indicate the decay of magnitude from 480 nm to 560 nm. This can also be confirmed by directly observations of the TA spectra in Fig.\ \ref{fig:Fig2}. The spectral profile of 5.84 ps in Fig.\ \ref{fig:Fig3}(d), however, shows quite different shape. It presents a positive and negative peaks at 500 and 520 nm, which imply a signature of population transfer from higher conduction band to lower band in InSe. In addition, the magnitude of this component is dramatically reduced. The decay associated component of 20.17 ps (green circles) shows a broad peak at 510 nm in Fig.\ \ref{fig:Fig3}(a), however, it gradually change the profile from (b) to (d). The positive and negative features of this component indicate the decay and grow of amplitude in TA spectra. Moreover, the component of 414 ps (light blue circles) shows a broadband from 500 nm to 560 nm. This component has been repeated from Fig.\ \ref{fig:Fig3}(b) to (d). The last component shows the spectral profile of residuals and plotted as pink circles from (a) to (d). To compare the spectral profiles of the time components that might represent hot carrier relaxation dynamics, we have overlayed the spectral components that we believe to represent the same physical processes after photexcitation with different pump excitations as Fig.\ \ref{fig:Fig4}. A spectrally narrowing of the ground state bleach features on reducing the pump excitation frequency can clearly be observed in Fig.\ \ref{fig:Fig4}(a). It is a very interesting observation, which will need further investigation to reveal the physical nature of this observed narrowing. Additionally, we observe reduction in the amplitude of the second time components on decrease of the pump excitation frequency, shown in Fig.\ \ref{fig:Fig4}(b). This can be attributed to the reduction in the excess kinetic energy given to the carriers on pump excitation near the edge of the bandgap. The overlay of the other time components representing recombination dynamics are provided in the supporting information.

\subsection{Vibrational coherences} 

\begin{figure}[h!]
\begin{center}
\includegraphics[width=17.0cm]{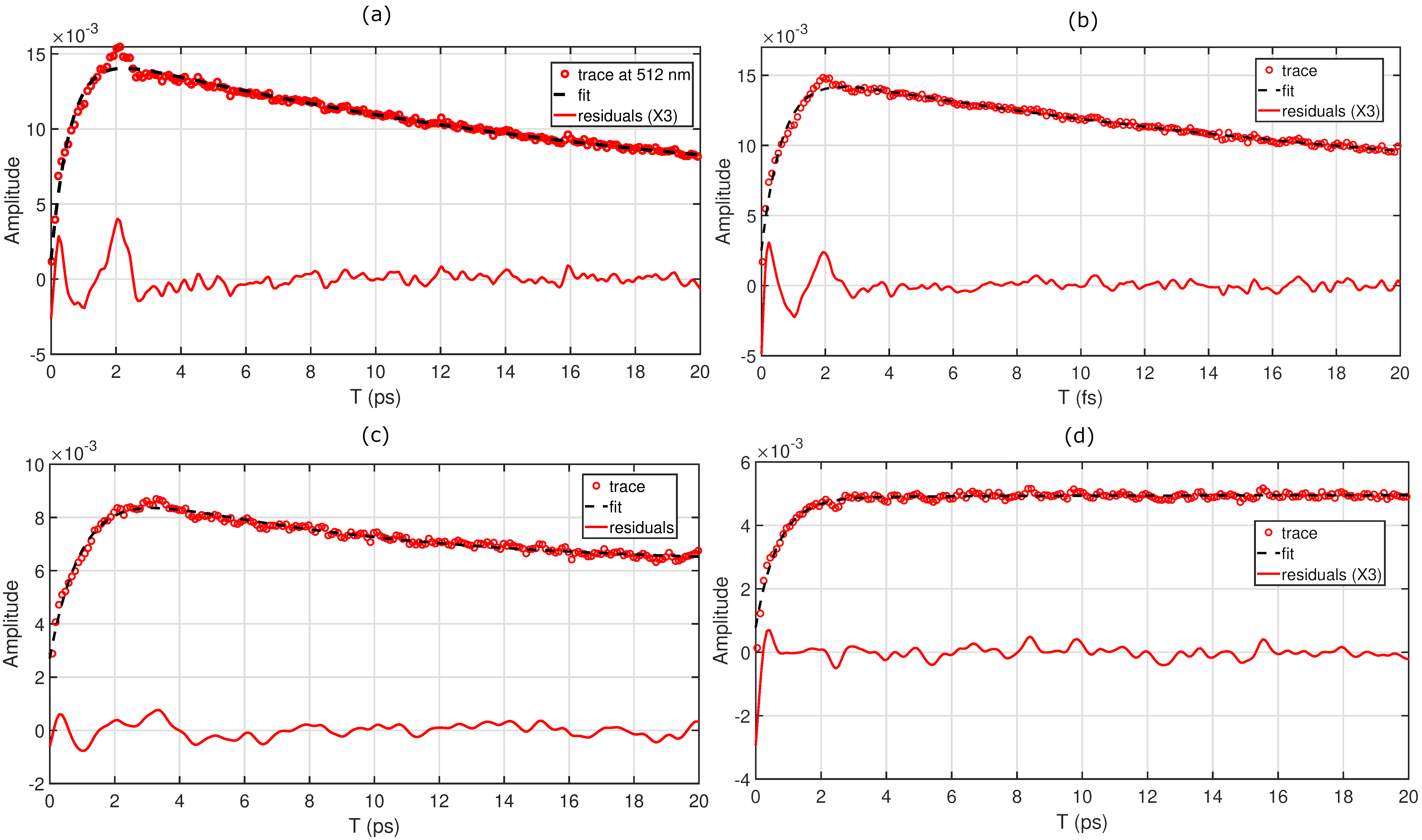}
\caption{\label{fig:Fig5} Time-resolved traces of GSB band (red circles) and the exponential fitting results (black dashed lines) for the transient absorption data obtained at pump excitation of 600, 650, 700 and 750 nm in (a), (b), (c) and (d), respectively. The residuals are shown as red solid lines and the magnitude of oscillations is magnified 3 times. } 
\end{center}
\end{figure}

\begin{figure}[h!]
\begin{center}
\includegraphics[width=17.0cm]{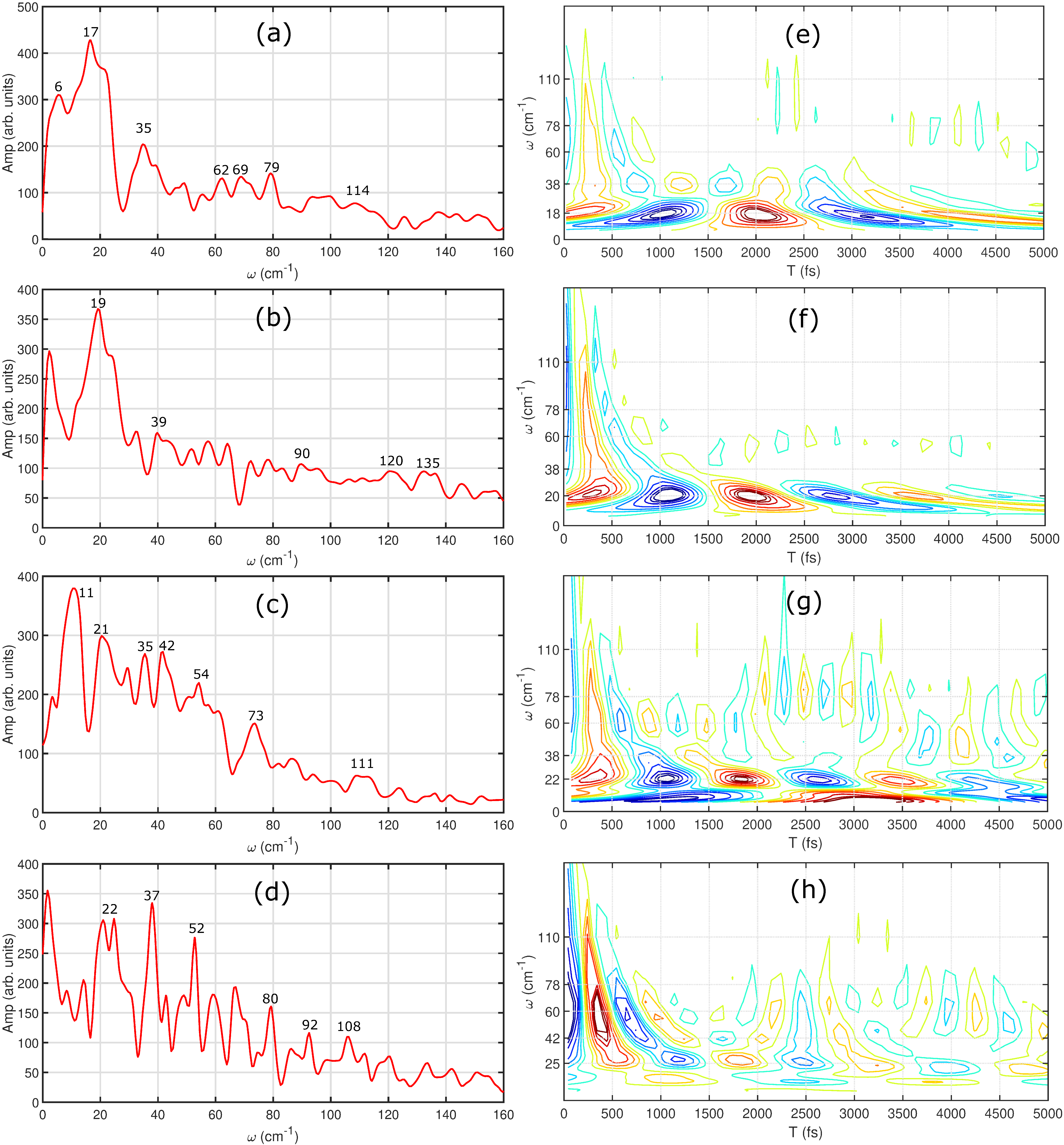} 
\caption{\label{fig:Fig6} Fourier transform of residuals and the results are shown in (a), (b), (c) and (d) for the excitation at 600, 650, 700 and 750 nm, respectively. The identified modes are marked as black numbers. The obtained results of associated wavelet analysis are shown from (e) to (h) with clarified frequencies of vibrational coherences.  } 
\end{center}
\end{figure}

To examine the coherent dynamics in InSe, we extract the kinetic profiles of GSB bands in TA spectra obtained at different pump excitation wavelengths, which are shown here in the Fig.\ \ref{fig:Fig5}. To improve the signal-to-noise, the plotted data are averaged over the kinetic profiles from 510 nm to 514 nm of the respective transient data, shown as red circles in Fig.\ \ref{fig:Fig5}. To retrieve the vibrational coherences convoluted with population dynamics in the kinetic data, we firstly perform the global fitting approach and show the fitting data of the GSB kinetic profiles as black dashed lines in Fig.\ \ref{fig:Fig5}. The obtained residuals from the kinetic fitting are shown as red solid lines in Fig. \ref{fig:Fig5}. The amplitude of residuals are magnified by 3 times to show the clear oscillatory dynamics. Then, we perform the Fourier transform of these residuals and the resulted vibrational modes are plotted as red solid lines in Fig.\ \ref{fig:Fig6} (a), (b), (c) and (d) for TA data at different pump excitations 600, 650, 700 and 750 nm, respectively. Many low-frequency modes at frequencies 18, 38, 60, 78 and 110 cm$^{-1}$ are prominently visible in the obtained vibrational power spectra. Intensity profiles of the observed vibrations varies for different pump excitation data. Assignment of the observed vibrational coherences have been described later in the discussion section. To examine the vibrational lifetimes of the observed normal modes, we employ the wavelet analysis on the residuals and plot the respective results from (e) to (h) in Fig.\ \ref{fig:Fig6}. The details of the wavelet analysis have been described in the SI. The results show that, in the low-frequency region, the vibrational mode at ($<20$cm$^{-1}$) has a clear oscillatory dynamics even after waiting time of 5 ps. Lifetime of these modes is dramatically reduced with the increase of vibrational frequencies. The mode at 38 cm$^{-1}$ decays within 2 ps and the frequency at $>60$ cm$^{-1}$ shows the lifetime shorter than 1 ps. 

\section*{Discussions} 

\begin{figure}[h!]
\begin{center}
\includegraphics[width=17.0cm]{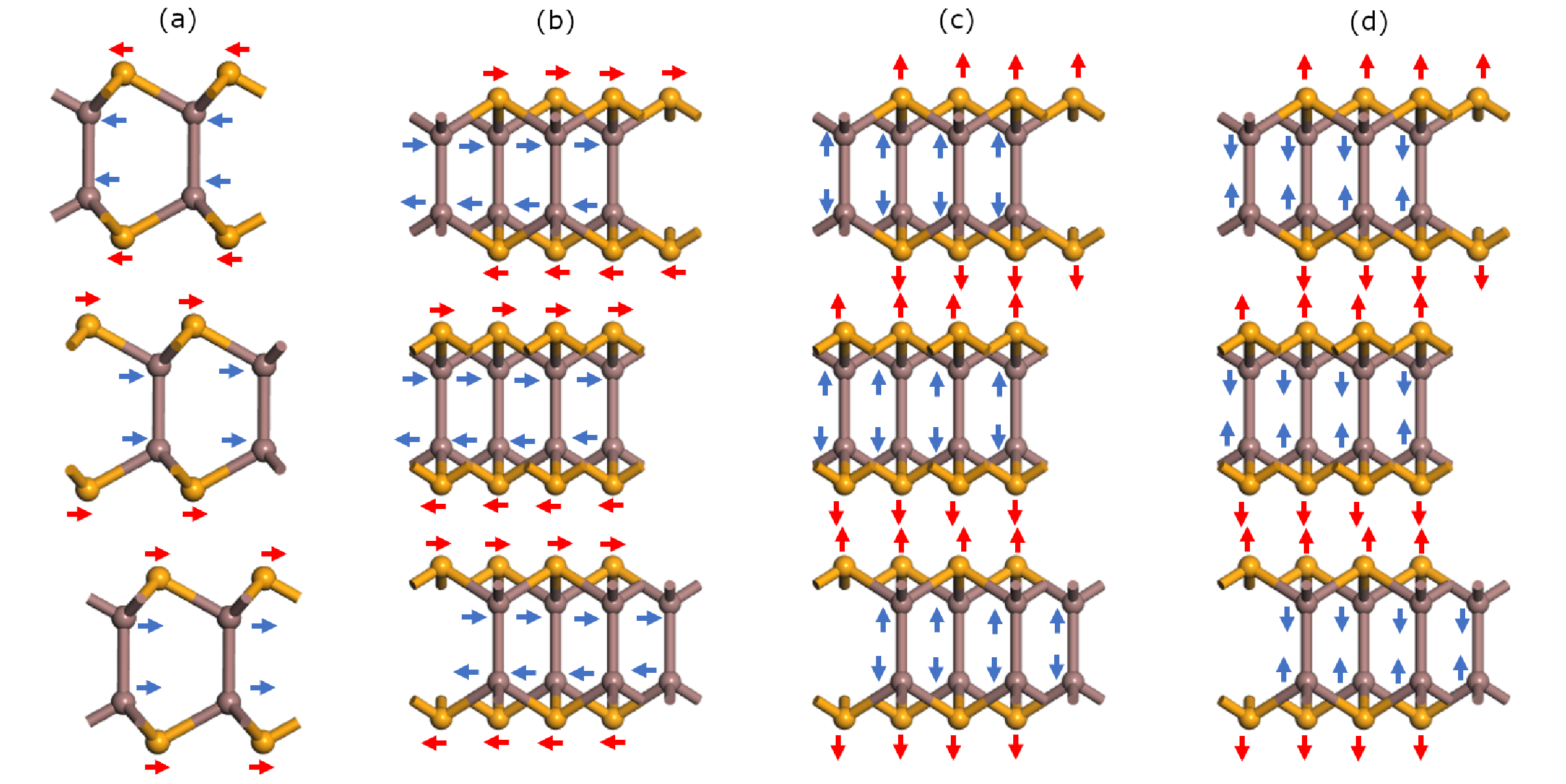}
\caption{\label{fig:Fig7} The calculated vibrational modes of 15 cm$^{-1}$, 38 cm$^{-1}$, 117 cm$^{-1}$ and 227 cm$^{-1}$ in (a), (b), (c) and (d), respectively. The motions of each atom in crystal are marked by the red and blue arrows.   } 
\end{center}
\end{figure}

Understanding the electron-phonon-coupling dynamics upon ultrafast photoexcitation of InSe is necessary for understanding its optoelectronic properties. The wavelength dependent data discussed earlier in the result sections presents information on how different phonons play role in carrier relaxation after photoexcitation from VB1 to CB. To retrieve the origin of these observed phonons in the spectroscopic measurements, we perform the ab-initio calculations of $\rm \gamma$-InSe. The details of theoretical calculations are presented in the Method section. We have presented the relevant vibrational modes at 15, 38, 117 and 227 cm$^{-1}$ in Fig.\ \ref{fig:Fig7}. These calculated vibrational frequencies agree to the observed vibrational coherences in our experimental data. Importantly, based on our vibrational analysis and associated wavelet analysis, we observe two timescales of phonon dynamics after photoexcitation: 1 ps and 5-10 ps. Our wavelet analysis reveal that the lifetime of vibrational coherences of a mode at ($<$20 cm$^{-1}$) is around 5 ps, while the mode of 38 cm$^{-1}$ decays within a timescale of 2 ps. Thus, we can infer that the carrier relaxation dynamics is mainly contributed by the interaction with the low-frequency phonons, which results into two different timescales of the experimentally observed relaxation timescales. So, the phonon mode at $<$20 cm$^{-1}$ with a lifetime of 5 ps may contribute towards the carrier relaxation dynamics observed at 5-10 ps, while the phonon modes with shorter lifetime may have contribution towards the earlier time relaxation processes. Theoretical calculations reveal that the motion of mode 15 and 38 cm$^{-1}$ strongly relate to the E'(1) and E''(1) motion between In and Se atoms. This particular motion of low-frequency modes is assigned to the intrelayer phonons \cite{JPCL 13 3691 2022}. Moreover, the vibrational coherences obtained for different pump excitation wavelengths show that the magnitude of the contribution of these phonons towards carrier relaxation hugely varies with initial kinetic energy of the carriers. Hence, the thermal energy redistribution of the charge carriers is dependent on the initial energy level that they are excited to. Additionally, it must be noted that the amplitudes of these wave-packet motions report on the magnitude of the excited-state structural changes. Hence, the structure-based dynamics of motion of each mode has been plotted in the Fig.\ \ref{fig:Fig7}, which can be used to interpret the initial photo-dynamics in an atomic level. 

\section*{Conclusions} 

We have studied the excitation wavelength dependence charge carrier dynamics in InSe single crystals using broadband femtosecond transient absorption spectroscopy. The excitation pump wavelength dependence near the band edge (A exciton; VB1 to CB transition) shows the faster cooling of the hot carriers with increasing pump wavelength. Additionally, our measurements show that the carrier thermalization process is biexpoential and varies from 1 to 10 ps with different excitations. This slow cooling has been explained to be caused by the large phononic band gap created by the large mass ratio between the anion (Se$^{2-}$) and cation (In$^{2+}$). The recombination shows two timescales: ~400 ps and $>$1 ns (beyond the available time-delays available in the current experimental setup). The recombination is reported to be primarily caused by the surface recombination \cite{JPCL 10 493 2019}. Our vibrational analysis of the observed coherences in the transient absorption data reveals the important low frequency lattice motions, which is coupled to electron relaxation dynamics. The nature of the long-lived phonons have been deciphered using ab-initio quantum calculations. Importantly, our mesurements distinctly captures short lived (~1 ps) and long-lived (~5 ps) vibrational coherences that might be contributing to the observed biexpoential thermal-relaxation dynamics occuring in the similar time-scales. It is a very unique observation since interactions of hot carriers to different phonons lead to differential thermalization dynamics in the system. Thus, our work provides a detailed mechanism of electrons interacting to the lattice phonons during relaxation dynamics after resonant photoexcitation of InSe. The detail photophysical understanding provided by this work will help to realize high-performance optoelectronic devices using InSe crystals. 

%


\section*{Materials and Methods}
\subsection{Sample preparation.} 

The bulk InSe crystal sample was purchased from 2D semiconductor (home page: www.2dsemiconductors.com) and used it without further modification. The thin slices of the InSe sample 50 nm have been prepared using microtome machine (Ultramicrotome Leica EM UC7). Prepared thin slides of samples have been directly attached to the quartz substrate and mounted on a XYZ translation stage, the pump/probe laser pulses are focused in the same spot on the surface of sample. 

\subsection{Femtosecond transient absorption measurements.} 

A detailed descrption of the optical setup is provided in an earlier reported work from our group \cite{SciRep 12 18216 2022}. Briefly, transient absorption measurements are performed in a Helios-EOS fire spectrometer from Ultrafast System. The primary femtosecond laser source is Astrella (800 nm, 100 fs, 7 mJ/pulse, and 1 kHz repetition rate, Coherent Inc.). To generate probe beam in visible region of the spectrum, a white light supercontinuum is generated by focusing on 3 mm thick sapphire crystal. The tunable pump beam is generated using a commercial OPA (OPerA Solo, Coherent Inc.). Following a 200 nm pump focus at the sample, the pump and probe pulses are spatially overlapped on the sample. For this system, a 120 fs instrument response function has been determined. The optical measurements have been performed after optimizing the measuring spot on sample surface. The XYZ translation stage has been modulated to find the best spot and reduce the scattering from the pump and probe pulses. The real measurements have been performed at room temperature after these refining procedures. The equally distributed time step of 200 fs has been used at initial waiting time up to 40 ps. The larger time step has been used for the rest of measurements up to 2 ns. MATLAB has been used to process the TA data using custom-written codes.

\subsection{Theoretical calculations.} 

We perform the theoretical calculations to examine the structural origin of observed vibrational modes. For this, we firstly use the crystal structure of $\gamma$-InSe from the X-ray diffraction measurement and the final structure used in the calculation has been refined according to literature (111-InSe.cif) \cite{PRB 48 14135 1993, IPPS 235 456 2003}. The density functional theoretical calculations for this study are performed by using the projector-augmented-wave (PAW) method and the Perdew-Burke-Ernzerhof (PBE) generalized gradient exchange approximation correlational functional implemented in Vienna ab initio simulation package (VASP) \cite{PRB 59 1758 1999, JCP 122 234102 2005, J Comput Chem 29 2044 2008}. For the initial geometrical optimization, the energy convergence criteria for electronic and ionic iterations are set to be 10$^{-4}$eV and 10$^{-5}$eV, respectively.  The energy cutoff for the plane-wave basis is set to 300 eV for all calculations. The Brillouin zone is represented by a Monkhorst–Pack special k-point mesh of 8$\times$8$\times$2 for the geometry optimization. To give an accurate explanation of the weak interactions such as London dispersion interaction and long-range van der Walls (vdW) intermolecular interactions, DFT-D3 method was applied \cite{Chem Phys Chem 12 3414 2011}. The phonon dispersion spectra and Raman spectral lines according to group theory are calculated based on the density functional perturbation theory (DFPT) method \cite{PRB 55 10355 1997}. In theory, the larger the supercell, the more accurate the dispersion curves that are obtained. We adopt the 2$\times$2$\times$2 rectangle supercell for all phonon calculations.

%
\begin{addendum}
\item This work was supported by the research grant of State Key Laboratory Spectroscopy, School of Physics and Electronic Science, East China Normal University and the starting grant of Ningbo University. The work was also supported by NSFC grant with NO.\ 12274247 and the foundation of national excellent young scientist. 

\item[Supporting information] The supplementary Information includes the global fitting approach and obtained results, the TA measurement with sample thickness and the details of wavelet analysis. The TA data measured with excitation at 400 nm. The TA data measured with different excitation energies. 

\item[Competing Interests] The authors declare that they have no competing financial interests. 

\item[Correspondence] Correspondence of paper should be addressed to A. J. (email: ajay.Jha@rfi.ac.uk), Y. Y. (email: yyang@lps.ecnu.edu.cn) and Z. R. S. (email: zrsun@phy.ecnu.edu.cn) 

\item[Author contributions] H. -G. D. and A. J. conceived the research and discussed with J. W. S. and Z. R. S.. J. W. S. and M. H. J.  performed the spectroscopic measurements. J. Y. L. H. T. S. and Q. H. Y. constructed the model and performed theoretical calculations. H. G. D. and A. J. wrote the initial draft and refined by all authors. Z. R. S. and A. J. supervised this project. 

\end{addendum}
%
\newpage

\newpage

\newpage

\newpage

\newpage

\newpage

\newpage

\end{document}